\documentclass[aps,prb,twocolumn,amssymb,floatfix]{revtex4}  
\usepackage{graphicx}

\usepackage{hyperref}

\begin{document}

\title{A combination of continuous-wave optical parametric oscillator and
femtosecond frequency comb for optical frequency metrology}
\author{E. V. Kovalchuk, T. Schuldt, and A. Peters}

\affiliation{Institut f\"{u}r Physik, Humboldt-Universit\"{a}t zu Berlin,
Hausvogteiplatz 5-7, 10117 Berlin, Germany}

\homepage{http://qom.physik.hu-berlin.de/}

\begin{abstract}
We combine a tunable continuous-wave optical parametric oscillator and a
femtosecond Ti:Sapphire laser frequency comb to provide a phase-coherent bridge
between the visible and mid-infrared spectral ranges. As a first demonstration
of this new technique we perform a direct frequency comparison between an
iodine stabilized Nd:YAG laser at 1064~nm and an infrared methane optical
frequency standard at $3.39\,\mu$m.\\
\begin{center}
Accepted for publication in \emph{Optics Letters} \copyright   2005 Optical
Society of America
\end{center}

\end{abstract}



\maketitle



\begin{figure}[b] 
  \centering
  \includegraphics[]{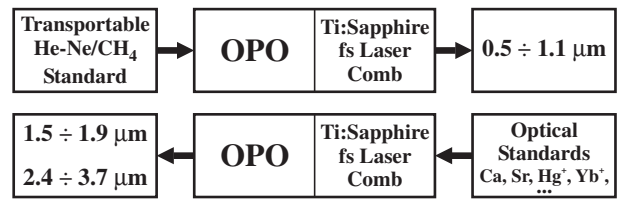}
  \caption{OPO-comb tandem as a bidirectional coherent link between IR and visible
  spectral ranges.}
  \label{OPOcomb}
  \end{figure}

\noindent Continuous-wave optical parametric oscillators (cw-OPOs) are one of
the most promising infrared (IR) laser sources for high-resolution molecular
spectroscopy.\cite{KovDek01, HeckHav03}  They offer wide tuning range and high
output power levels,\cite{StrMeyn02,StoLin04,HerLi02} as well as narrow
linewidth and the possibility of phase locking to stable optical
references.\cite{KovDek02, InaIke04} Combining them with recently developed
optical frequency combs based on femtosecond
lasers\cite{JonDid00,UdemHol02,TelSte99,DidJon00} opens up new perspectives in
optical metrology. In such a system  the special properties of the OPO allow it
to serve as a bidirectional coherent bridge linking the IR and visible spectral
ranges (Fig.~\ref{OPOcomb}).\cite{KovDek02} For example, an infrared methane
optical frequency standard\cite{GubTyu95,GubShel99} can be used as a highly
stable reference for an OPO and a visible frequency comb. Alternatively, an OPO
phase locked to a visible frequency comb, which itself is referenced to a high
performance microwave or optical clock, can provide stable emission with known
absolute frequency in both the mid-IR and telecom spectral ranges. In this
Letter we present the first implementation of this new technique.

Methods of femtosecond optical frequency metrology are based on the
establishment of a phase-coherent relation between an optical frequency and
that of a primary microwave clock or of an optical frequency standard. This is
achieved using a comb of equidistant lines with frequencies
$f_n=n\,f_{rep}+f_0$, where $n$ is an integer, $f_{rep}$, the repetition rate,
and $f_0$, the carrier-envelope offset frequency of the femtosecond laser, with
$f_0<f_{rep}$.\cite{JonDid00,UdemHol02,TelSte99,DidJon00} Current optical
frequency combs are predominantly based on mode-locked Ti:Sapphire and Er:fiber
lasers. Their output is thus restricted to the visible and near-IR spectral
ranges, while building an optical clockwork capable of addressing the mid-IR
range has remained a challenge. Application of frequency combs for this purpose
requires their combination with additional steps of sum- or
difference-frequency generation (SFG and
DFG).\cite{AmyGon04,MueKuz04,ForJon03,ZimGoh04,ForMar05}

SFG clockworks need for their realization additional transfer oscillators such
as diode lasers or a commercial cw-OPO.\cite{AmyGon04,MueKuz04} Offset-free DFG
combs, which are produced in the IR by mixing either the output of two
synchronized visible frequency combs\cite{ForJon03} or different spectral
components of a specially designed Ti:Sapphire laser,\cite{ForMar05} have very
low output power (tens of $\mu$W total, with $<1\,$nW per comb mode). All these
approaches phase-coherently connect IR frequencies and primary microwave clocks
via $f_{rep}$ but not the optical frequencies. They are either tailored to very
specific goals or quite complex, and generally they require a number of
non-linear crystals and supplementary lasers to make them useful for
spectroscopic applications. Thus, a specific benefit of the new method
presented here is that it consolidates all these subsystems into a single
cw-OPO, providing direct phase-coherent link between two optical frequencies
--- the visible and the infrared.

 \begin{figure}[t] 
  \centering
  \includegraphics[]{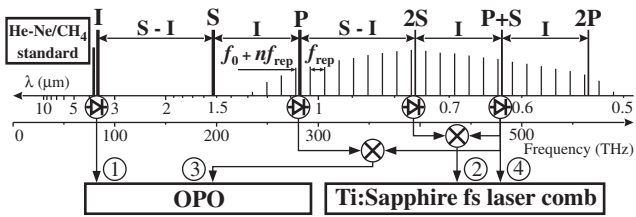}
  \caption{Scheme of phase locking of the OPO output frequencies and a Ti:Sapphire femtosecond
  laser comb to a He--Ne/CH$_4$ frequency standard.}
  \label{OPOcomblock}
  \end{figure}

%
%
%
%
The scheme itself is based on the fact that a singly resonant + pump resonant
cw-OPO emits not only strong signal (S) and idler (I) waves, but also a set of
weak (a few $\mu$W) components resulting from non-phase-matched mixing
processes of the resonated pump (P=S+I) and signal waves.\cite{KovDek02} Some
of these (2S, P+S, 2P) are located within the emission range of a Ti:Sapphire
femtosecond laser comb. Forming suitable differences of the heterodyne beat
frequencies with adjacent comb lines then allows mutual phase locking of OPO
optical frequencies and the microwave frequencies $f_{rep}$ and $f_0$
characterizing the femtosecond comb. This can be implemented following a
variety of schemes, e.g. the basic four-step locking scheme as illustrated in
Fig.~\ref{OPOcomblock}: \\[3 mm]
\noindent 1.  Phase lock the OPO idler frequency I to a He--Ne/CH$_4$ standard. \\[1.5 mm]
2.  Phase lock the comb spacing $f_{rep}$ to the frequency difference between
the lines P+S and 2S, which is equal to the already stabilized idler frequency
I, [(P+S)-2S=I]. \\[1.5 mm]
3.  Measure and stabilize the difference between frequencies (P+S) and P, and
therefore the signal frequency S using the comb, [(P+S)-P=S]. As a result, the
stability of the methane standard is transferred to the signal
and pump frequencies, and thus the entire OPO spectrum. \\[1.5 mm]
4.  Measure and stabilize the comb offset frequency $f_0$ relative to one of
the OPO lines, e.g. P.\\[3 mm]
The first two steps by themselves are already sufficient to implement a
methane-based infrared optical molecular clock analogous to that reported by
Foreman {\em et al}.\cite{ForMar05} The last two steps then phase lock the
whole femtosecond frequency comb and all OPO output lines to the He--Ne/CH$_4$
standard.

As a first realization of this method we have performed a direct comparison
between a Nd:YAG laser (1064~nm) stabilized on the a$_{10}$ line of the R(56)
32--0 iodine transition near 532~nm\cite{MueHer03} and a He--Ne laser
stabilized on the $\sim$300~kHz wide $P(7)\,$F$_2^{(2)}$ line of the methane
molecule at 3.39~$\mu$m. The latter is integrated in a transportable
He--Ne/CH$_4$ optical frequency standard and serves us as a highly stable (over
hundreds of seconds) IR reference laser\cite{GubTyu95}. The methane standard
and reference laser were both previously characterized during several absolute
frequency comparisons.\cite{GubShel99}

\begin{figure}[b]  
  \centering
  \includegraphics[]{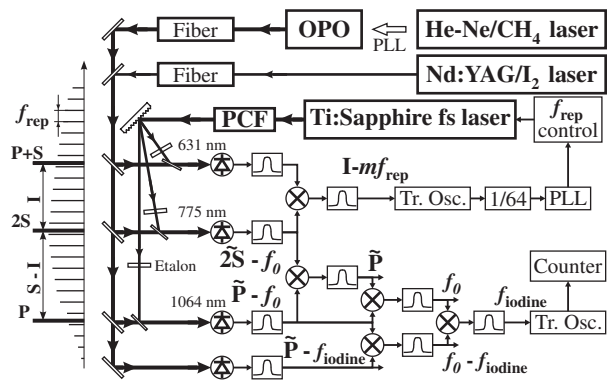}
  \caption{Diagram of the frequency comparison I$_2$ vs CH$_4$. PCF, photonic crystal
  fiber; Tr. Osc., tracking oscillator; $m$ is an integer.
   $\widetilde{P}$ and $\widetilde{2S}$ are radio frequencies exactly mirroring
   changes in P and 2S.}
  \label{ComparisonSetup}
  \end{figure}

The experimental arrangement of the frequency comparison is shown in
Fig.~\ref{ComparisonSetup}. The cw-OPO was specifically developed for
applications in high-resolution Doppler-free molecular spectroscopy and
metrology.\cite{KovDek01,KovDek02} It is similar to a system described
earlier,\cite{KovDek01} though modified to be mechanically more stable and
featuring improved cavity lock electronics. The setup is based on a
periodically-poled lithium niobate (PPLN) crystal with multiple grating
periods, pumped by a monolithic 1-W Nd:YAG laser. The pump and the signal waves
are both resonated in the same, folded cavity, the length of which is locked to
the pump laser using a piezoelectric transducer (PZT). A specially designed
temperature-stabilized intracavity etalon allows controlled access to any
desired wavelength in a wide OPO emission range: 1.5--1.9~$\mu$m (signal) and
2.4--3.7~$\mu$m (idler). Using this configuration, we obtain well-defined
tuning behavior of the idler output radiation at power levels of $>$50~mW at
3.39~$\mu$m, exceptional long-term stability, and an instantaneous linewidth of
about 10~kHz. The idler frequency can easily be phase locked to any optical
reference using a phase-locked loop (PLL) with a bandwidth of $\sim$15~kHz and
by applying the correction signal to the pump laser PZT.\cite{KovDek02}

Our frequency comb is based on a femtosecond Ti:Sapphire ring laser
(GigaOptics, GigaJet-20) with a repetition rate $f_{rep}\sim$750~MHz. A
continuum between 0.5 and 1.1~$\mu$m with a total power of $\sim$300~mW is
generated in a photonic crystal fiber and then split with a diffraction grating
into three spectral parts, centered around the 1064~nm, 775~nm and 631~nm ---
corresponding to the OPO lines P, 2S and P+S. After additional filtering with
Fabry-Perot etalons these beams are overlapped with the related OPO components
and sent to avalanche photodiodes. Typical power levels here are 10, 0.3 and
1~$\mu$W  for P, 2S and P+S, respectively. The three resulting OPO beat signals
with adjacent comb lines typically have signal-to-noise ratios of 25-40~dB in a
resolution bandwidth of 100~kHz. In a slight modification of the more general
scheme presented above we measured the iodine frequency relative to the OPO
pump laser.

During the comparisons only $f_{rep}$ was phase locked to the methane standard
while the comb offset $f_0$ remained free-running, similarly to the approach of
Telle {\em et al}.\cite{TelLip02} Frequency differences between the relevant
filtered and amplified beats were processed using double-balanced mixers.
Frequency generators and counters involved were referenced to a Rubidium
microwave standard. The signal used for phase locking of the comb spacing to
the stabilized idler frequency and the final signal corresponding to the iodine
stabilized laser frequency were both refined using tracking oscillators. The
iodine frequency signal was then recorded using a PC-based counter board.

\begin{figure}[t] 
  \centering
  \includegraphics[]{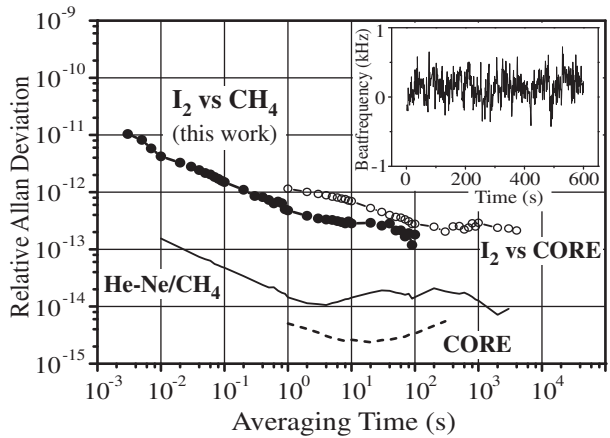}
  \caption{Result of the frequency comparison between the iodine and methane
  stabilized lasers.
  Also shown are independent measurements of a similar iodine
  system\protect\cite{MueHer03} with respect to a cryogenic optical
  resonator (CORE) (open circles) and known performances of the methane
  standard\protect\cite{GubShel99} (solid line) and the CORE\protect\cite{SeeSto97} (dashed
  line). Inset: typical beat signal of the I$_2$ vs CH$_4$ comparison,
  counted with a 1~s gate time.}
  \label{AViod}
  \end{figure}

Fig.~\ref{AViod} shows the resulting relative Allan deviation averaged over
different measurement runs (counter gate times between 1~ms to 10~s). The
result of the comparison is limited by the iodine stabilized laser performance,
which is known from independent measurements relative to a cryogenic optical
resonator (CORE).\cite{MueHer03} The accuracy limitation of the new comparison
method itself is expected to be much lower than $10^{-13}$. Continuous
measurement time was restricted by the slow degradation of the photonic crystal
fiber, which led to non-uniformities in the comb spectrum and insufficient
power levels. The OPO showed very reliable operation and remained phase locked
to the methane stabilized laser over several days. We also succeeded in
reversing the scheme and phase locking $f_{rep}$ and the idler frequency to the
iodine stabilized laser.

As the current comparisons do not yet challenge the accuracy performance of the
system, upcoming measurements will use better optical frequency references
(e.g. CH$_4$ vs CORE) to explore the actual limitations. The ease-of-use of the
setup could be improved by using specially designed PPLN crystals in order to
enhance the power of the non-phase-matched OPO components 2S and P+S. Another
line of development is to use the opposite direction of stability transfer in
order to provide stable emission for the purposes of high-resolution
Doppler-free molecular spectroscopy in the infrared with output power levels up
to several Watts using a modified design.\cite{StrMeyn02,HerLi02} Specifically,
this should lead to applications in precision IR spectroscopy, such as
metrology of the strong transitions of cold CH$_4$ molecules\cite{KovDek02} and
the study of rotation-vibrational transitions in decelerated and trapped
OH-radicals.\cite{MeeSme05}

We are very grateful for the inestimable support by J. Mlynek and wish to thank
J. Knight (University of Bath, UK) for providing the photonic crystal fiber and
A. Bauch (PTB Braunschweig, Germany) for making available the Rubidium
standard. E. Kovalchuk (evgeny.kovalchuk@physik.hu-berlin.de) appreciates
support from G. Ertl and G. Meijer (Fritz-Haber Institute of the MPG, Germany).
He is also with the Frequency Standards Laboratory at P. N.
Lebedev Physics Institute, Moscow, Russia.  \\
T. Schuldt is also with EADS Astrium GmbH, 88039 Friedrichshafen, Germany.



\end{document}